# Governing Information Security in Conjunction with COBIT and ISO 27001

Tolga MATARACIOGLU[1] and Sevgi OZKAN[2]

[1]TUBITAK National Research Institute of Electronics and Cryptology (UEKAE),
Department of Information Systems Security, 06700, Ankara, TURKEY
[2]Middle East Technical University, Informatics Institute,
Department of Information Systems, 06531, Ankara, TURKEY

mataracioglu@uekae.tubitak.gov.tr, sozkan@ii.metu.edu.tr

## ABSTRACT

*In this paper, after giving a brief definition of Information Security Management Systems (ISMS), ISO 27001, IT governance and COBIT, pros and cons of implementing only COBIT, implementing only IS0 27001 and implementing both COBIT and ISO 27001 together when governing information security in enterprises will be issued.*

## I. INTRODUCTION

Information Security Management System (ISMS) is a set of processes and the main goal of those systems is to manage information security issues in an enterprise [6]. ISMS uses Plan-Do-Check-Act (PDCA) model and the input of this model is information security requirements and expectations. The expected output is obviously managed information security. In Plan phase, establishing ISMS policy, objectives, processes and procedures relevant to managing risk and improving information security to deliver results in accordance with an organization's overall policies and objectives are performed. In Do phase, implementation and operation of the ISMS policy, controls, processes and procedures are performed. In Check phase, assessment and, measurement of process performance against ISMS policy, objectives and practical experience and reporting of the results to management for review are performed. In Act phase, taking corrective and preventive actions, based on the results of the internal ISMS audit and management review or other relevant information, to achieve continual improvement of the ISMS is performed.

ISO 27001 is the standard of implementing ISMS to an enterprise. To implement each of the phase of the PDCA model, items numbered 4.2.1, 4.2.2, 4.2.3 and 4.2.4 are given in the standard. The standard has ten domains including security policy, organizational security, asset classification and control, access control, compliance, personnel security, physical and enviromental security, system development and maintenance, communications and operations management and business continuity management.

Governance requires a balance between the conformance (i.e. adhering to legislation, internal policies and audit requirements) and performance (i.e. improving profitability, efficiency, effectiveness and growth) goals, as directed by the board [7]. IT (information and related technology) governance is defined as a structure of relationships and processes to direct and control the enterprise toward achieving its goals by adding value while balancing risk versus return over IT and its processes [7].

The best practice of implementing IT governance is COBIT (Control Objectives for Information and Related Technology). According to COBIT, principles of IT governance are direct and control, responsibility, accountability and activities. Also the focus areas are given as strategic alignment, value delivery, risk management, resource management and performance measurement. The delivery of information is controlled through 34 high-level objectives, one for each process. For controlling this delivery, COBIT provides three key components, each forming a dimension of the COBIT cube: Business requirements, IT resources and IT processes. COBIT has 4 domains [7].

In Plan and Organize (PO) domain, formulating strategy and tactics, identifying how IT can best contribute to achieving business objectives and planning, communicating and managing the realization of the strategic vision are performed. This domain consists of 10 processes. In Acquire and Implement (AI) domain, changing and maintaining existing systems and identifying, developing or acquiring, implementing and integrating IT solutions are performed. This domain consists of 7 processes. In Deliver and Support (DS) domain, service support for uers, and the management of security, continuity, data and operational facilities are performed. This domain has 13 processes. In Monitor and Evaluate (ME) domain, performance management, monitoring of internal control, regulatory compliance and governance issues are performed. This domain consists of 4 processes. The business requirements are effectiveness, efficiency, confidentiality, integrity, availability, compliance and reliability. And IT resources are applications, information, infrastructure and people according to COBIT.

After giving the objectives of this paper in Part II, pros and cons of implementing only COBIT, implementing only IS0 27001 and implementing both COBIT and ISO 27001 together when governing information security in enterprises will be issued in Part III. The conclusion and the results of the paper will be given in Part IV.

## II. OBJECTIVE

The main objective of the paper is to relate and construct a mapping between COBIT framework and ISO 27001 standard when governing an enterprise. Both of the frameworks are complementary and may be more beneficial to enterprises provided that they are used together to fulfill the information security governance issues.

So as to govern an enterprise fully, integration of COBIT and ISO 27001 issues is indispensable. Implementing only COBIT addresses all of the information security duties. However, several standards like ISO 27001, describe the duties in a more comprehensive manner than does COBIT. Thus, in order to implement the governance in the enterprises, other standards like ISO 27001 have to be considered.

## III. DISCUSSION

Implementation of ISO 27001 in order to manage the security of an enterprise has some advantages. ISO 27001 certification serves as a public statement of an organization's ability to manage information security [2]. It ensures that its information security management system and security policies continue to evolve and adapt to changing risk exposures. Furher, these organizations will spend less money recovering from security incidents, which may also translate into lower insurance premiums [2] [4]. Also this standard is more detailed than COBIT, and provides much more guidance on precisely "how" things must be done [1].

Also ISO 27001 has some disadvantages when implemented alone in order to manage information security. It is a stand alone guidance and it is not integrated into a wider framework for IT governance.

IT governance has some benefits. Some of those are more reliable services, more transparency, responsiveness of IT to business, confidence of the top management and higher return on investment [7].
Some advantages of COBIT are given below [7]:

1. COBIT is aligned with other standards and best practices and should be used together with them.
2. It's framework and supporting best practices provide a well-managed and flexible IT environment in an organization.

3. COBIT provides a control environment that is responsive to business needs and serves management and audit functions in terms of their control responsibilities.
4. It provides tools to help manage IT activities.

The downside of using COBIT for IT governance is that it is not always very detailed in terms of "how" to do certain things. The control objectives are more addressed to the "what" must be done.

It therefore seems logical that to get the benefits of both the wider reference and integrated platform provided by COBIT, and the more detailed guidelines provided by ISO 27001, there can be a lot of benefit in using both together for information security governance [1].

Information Society Strategy 2006-2010 Activity Plan, prepared by T.R. Prime Ministry State Planning Organization, consists of several items including item number 88. This item identifies National Information Systems Security Programme. In this scope, ISO 27001:2005 based ISMS establishment consultancy is performed in four public bodies in Turkey by TUBITAK UEKAE. However, since there does not exist an IT governance awareness in those public bodies, benefits of establishing ISMS have not been seen. Some of the reasons are given below [8]:

1. TUBITAK UEKAE couldn't find a chance to get into touch with the board of the two of the public bodies.
2. Private personnel allocation could not be performed by the public bodies except one.
3. The allocated personnel have spent to ISMS establishment only a couple of his work hours in a week.
4. Establishment of ISMS has been tightened only within IT department.

Some of the misperceptions by public body boards and personnel are given below [8]:

1. Scope of the ISMS is IT department.
2. The responsible of ISMS establishment is the head of IT department.
3. ISMS is an information technology process.
4. Establishment of ISMS can thoroughly be done by other organizations.

The standard sentences to those misperceptions must be as given below [8]:

1. Scope of the ISMS is consequently the whole organization.
2. The responsible of ISMS establishment is the head of the organization.
3. ISMS is not an information technology process, indeed it is an information security process.
4. Consultancy service procurement can be done, however the main organization that has to establish ISMS is the organization itself.

So as to establish an ISMS to an organization, IT governance awareness should be complete among the organization. So ISMS and IT governance, or ISO 27001 and COBIT is highly related to each other. When an organization wants to establish ISMS and get ISO 27001 certificate, it has to take care about the issues that COBIT says, and vice versa. There also exists a mapping between COBIT and ISO 27001 in [3] [5]. It is a kind of building a bridge between COBIT and ISO 27001. The key point is to govern information security not only using ISO 27001 or COBIT alone, but in conjunction with those two in an enterprise.

## IV. CONCLUSION

In this paper, after giving a brief definition of Information Security Management Systems (ISMS), ISO 27001, IT governance and COBIT, pros and cons of implementing only COBIT, implementing only ISO 27001 and implementing both COBIT and ISO 27001 together when governing information security in enterprises have been addressed.

In Introduction part, definitions of ISMS, ISO 27001, IT governance and COBIT have been given. Also the domains and the components of ISO 27001 standard and COBIT framework have been summarized.

After giving the objectives of this paper in Part II, pros and cons of implementing only COBIT, implementing only IS0 27001 and implementing both COBIT and ISO 27001 together when governing information security in enterprises have been addressed in Part III, Discussion. Also some cases that TUBITAK UEKAE has owned, have been shared.

Implementing only ISO 27001 has some advantages in governing information security. Also COBIT has some advantages, too. However, both of those standards or best practices have some disadvantages when using alone. In order to camouflage the cons of ISO 27001, COBIT has to be taken into account. Further, so as to camouflage the disadvantages of COBIT as indicated above, ISO 27001 has to be taken into account, too. As a result, ISO 27001 and COBIT has to work together in providing the information security governance.

The mapping tool indicated in [5] gives some benefits to the organizations who has to govern their information security in conjunction with COBIT and ISO 27001. Linking corporate governance more directly to information security, senior management will have to face his responsibility more directly. Because corporate governance includes the responsibility for solid internal control, and internal controls rely on information security, information security is an integral part of corporate governance [4].

Only COBIT addresses the full spectrum of IT governance duties, however several standards describe the duties in a more comprehensive manner than does COBIT [3]. Thus, when implementing IT governance, those standards have to be considered and the guidelines, models and processes should be used to facilitate the implementation of COBIT [3].

If not, i.e. not using COBIT in establishing ISMS, the organizations will fail to get all of the benefits of ISMS, and vice versa.

**Tolga MATARACIOGLU**

After receiving his BSc degree in Electronics and Communications Engineering from Istanbul Technical University in 2002 with high honors, he received his MSc degree in Telecommunications Engineering from the same university in 2006. He is now pursuing his PhD degree in Information Systems from Middle East Technical University. He is working for TUBITAK National Research Institute of Electronics and Cryptology (UEKAE) at the Department of Information Systems Security as senior researcher. He is the author of many papers about information security published nationally and internationally. He also trains various organizations about information security. His areas of specialization are system design and security, operating systems security, and social engineering.

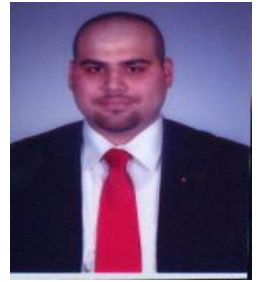

**Sevgi OZKAN**

Dr Sevgi Ozkan is an Assistant Professor at Department of Information Systems, Informatics Institute, Middle East Technical University Turkey. She is currently the Associate Dean of the School. She holds a BA and an MA in Engineering from Cambridge University and an MSc in Business Information Systems London University UK. She has a PhD in Information Systems Evaluation. She is a Fellow of the UK Higher Education and a Research Fellow of Brunel University UK. Since 2006, Dr. Ozkan has been involved with a number of EU 7th Framework and National projects in e-government.

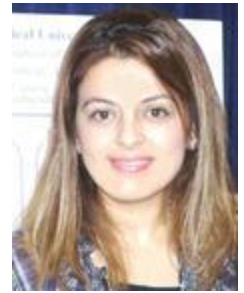

___________________________